\documentclass[12pt]{article}
\usepackage[utf8]{inputenc}
\usepackage[a4paper,margin=1in]{geometry}
\usepackage{amsmath,amssymb,bm}
\usepackage{tikz}
\usepackage{graphicx}
\usepackage{setspace}        
\usepackage{hyperref}
\title{Microscopic Pathways to Helix Formation: Packing Stabilization and Sticky Interactions in Chiral Polymer Condensates}
\author{\textbf{Biman Bagchi}\\
\small Solid State and Structural Chemistry Unit,\\
\small Indian Institute of Science, Bengaluru 560012, India}
\date{}
\begin{document}
\maketitle

\begin{abstract}
Helices are not generic outcomes of polymer collapse.
Collapsed conformations of semiflexible polymers with isotropic attractions generally form globules, toroids, or rod-like structures, as observed in simulations and described by coarse-grained necklace and surface-tension models. Helical conformations, by contrast, are generically absent in minimal theories based solely on bending elasticity and isotropic cohesion, because such descriptions provide no mechanism that selects torsion, pitch, or periodic packing.
Here we identify two minimal and physically distinct routes by which helices can become stable without invoking biochemical specificity. Route (A) is geometric and steric in origin: combining a tube-like packing (thickness) constraint with generic attractions selects an ideal helical packing with finite radius and pitch. Left- and right-handed helices remain exactly degenerate in free energy, so that chirality emerges spontaneously even in the complete absence of explicit chiral interactions.
Route (B) is energetic and commensurate: periodic “sticker” attractions between monomers separated by a fixed contour distance 
$m$ enforce a registry between interaction spacing and chain geometry. This commensurability selectively stabilizes helical states by allowing the same set of monomers to form repeated contacts along the backbone, naturally connecting the theory to classical Gibbs--DiMarzio and Zimm--Bragg mechanisms relevant for biopolymers.
For both routes, we derive analytical relations for the helix radius and pitch, the associated curvature and bending energy, contact-distance constraints, and crossover conditions to toroidal and rod-like morphologies, expressed in terms of the persistence length $L_{p}$, 
interaction strength, and chain length $N$. The theory thus explains why helices are non-generic in polymer collapse, identifies the physical ingredients required to stabilize them, and yields testable predictions for when helical and chiral condensates should appear.

\end{abstract}


\section{Introduction and motivation}

The seminal papers of Pauling and coworkers established the role of hydrogen bonding in the 
formation of helices and beta sheets in proteins and DNA \cite {Pauling1951, PaulingCorey1951,PaulingDNA}
Pauling had earlier pointed out the role of hydrogen bonding in biology\cite{Pauling1936}. 
Subsequent studies showed that helices are omnipresent. 
Yet, the role of hydrogen bonding and non-bonding interactions such as packing in formation of helices have
not been fully explored in theoretical studies. The present article is an effort to remove this lacuna.

The collapse of a single polymer chain in a poor solvent is a classic problem in polymer
physics and statistical mechanics. Minimal theoretical descriptions based on entropic
elasticity, short-ranged isotropic attractions, surface tension, and bending rigidity predict
a sequence of conformational morphologies such as globules, toroids, and rodlike states,
depending on solvent quality, chain stiffness, and molecular weight \cite {Flory1953,DeGennes1979,GrosbergKhokhlov1994,Lifshitz1978,
Odijk1986,KhokhlovSemenov1981}. These morphologies are now well understood within a
unified free-energy framework and form the backbone of modern theories of single-chain
collapse. In particular, the competition between cohesive driving forces (often represented
as a surface-tension penalty associated with polymer--solvent interfaces) and elastic
penalties (bending for semiflexible chains, stretching for finite extensibility) provides a
systematic route to morphology diagrams in which globules, toroids, and rods occupy large
and physically intuitive regions of parameter space.Thermal fluctuations play a central role in determining the conformational
statistics and dynamics of polymer chains.  The interplay between
elasticity, hydrodynamic interactions, and stochastic fluctuations,
discussed extensively in the classic monograph of Doi and Edwards
\cite{DoiEdwards1986}, provides the theoretical foundation for
understanding how polymer chains explore a wide range of morphologies
in solution.

Condensation of long semiflexible polymers into ordered structures has
been observed experimentally in systems such as DNA, where multivalent
counterions drive collapse into compact toroidal and rodlike aggregates
\cite{Bloomfield1997}.  Theoretical analyses have shown that electrostatic
interactions, bending elasticity, and packing constraints together
determine the geometry and stability of these condensates, including the
well-known toroidal DNA structures studied by Ubbink and Odijk
\cite{UbbinkOdijk1999}.

It is tempting to imagine that a helix should appear naturally as simply
another ``ordered'' way of packing a chain in three dimensions.
Yet both theory and simulation show that this expectation is generally
incorrect. Extensive molecular simulations of semiflexible polymers,
including bead--spring models based on the widely used Kremer--Grest
framework, show that collapsed chains typically form globules, toroids,
or rodlike structures rather than helices
\cite{KremerGrest1990,GrosbergKhokhlov1982,NoguchiYoshikawa1998,SchnurrMacKintosh2000}.
These studies demonstrate that the standard physical ingredients of
polymer collapse—excluded volume, chain stiffness, and isotropic
attractive interactions—are sufficient to generate ordered condensed
morphologies but do not generically stabilize helical structures.
The simulations of Srinivas and Bagchi are especially instructive in this
regard: they demonstrate, in a controlled and physically transparent
setting, that the same minimal ingredients (semiflexibility, isotropic
attraction, and thermal fluctuations) robustly produce globular and
ordered collapsed morphologies such as toroids and rods, yet do not
yield a stable helix \cite{SrinivasBagchi2002}.

This conclusion is reinforced at the analytical level. Helices are also absent from the
minimal surface-tension-plus-bending-energy theory. The reason is fundamental: a helix
necessarily carries both curvature and torsion. While curvature is penalized by bending
elasticity, torsion is neither rewarded nor selected in models with purely isotropic
attractions. In a purely ``surface + bend'' theory, the helix pays a curvature cost
\emph{uniformly along its entire contour} without receiving any compensating energetic gain
associated with twisting or with periodic registry of contacts. Competing morphologies can
avoid this handicap: a rod can largely eliminate curvature, while a toroid confines curvature
to a well-defined radius that can be optimized jointly with surface contributions. Thus, in
the absence of an additional physical mechanism that favors periodic registry, packing
regularity, or commensurate contacts, helices are generically unstable relative to toroids
or rods.

At the same time, ordered cylindrical and helical-like collapsed conformations \emph{do}
appear in specific physical settings, indicating that the instability of helices is not an
absolute prohibition but rather a statement about minimal models. A particularly relevant
example is provided by the experimental and simulation study of Hu \emph{et al.}, who showed
that stiff conjugated polymers with chemical defects can collapse into ordered cylindrical
conformations rather than featureless globules \cite{HuBarbaraRosskyBagchi2000}. 

There had been interesting studies on the relationship between global curvature, thickness and resulting shapes
of strings and helices. Helices are interesting geometric objects in their own interest \cite{GonzalezMaddoks1999,
MaritanHelix2000,BanavarMaritan2003}.

The importance of this work for the present paper is twofold. First, it provides a concrete and
high-visibility demonstration that ordered (cylindrical or helical-like) condensed states
can be realized for single polymer chains when additional structural or interaction scales
are present. Second, it motivates the central theoretical question we address here:
\emph{what are the minimal, generic mechanisms that can stabilize a helical condensate,
beyond the standard surface-tension and bending ingredients that already explain globules,
toroids, and rods?}

The aim of the present work is therefore to identify and analyze \emph{minimal} mechanisms
by which helical condensates can become thermodynamically stable without invoking
biochemical specificity, directional hydrogen bonding, or explicit chiral interactions.
We deliberately restrict attention to generic polymeric ingredients and ask a basic
question: under what conditions can a single polymer chain form a stable helix, and how
can such a state be described analytically at the free-energy level? Our focus is not on
local helix--coil ordering of a backbone segment, but on a \emph{global condensate
morphology} of an entire chain—a geometry that must compete, at the level of free energy,
against toroids and rods that are already favored by the minimal theory.

The theory we develop examines two complementary routes for helix stabilization. \\
(i) In Route~A, the polymer is treated as a tube of finite effective thickness. Steric
constraints then impose both a local curvature bound and a nonlocal self-avoidance
condition. When combined with generic attractions favoring dense packing, these constraints
select a helical geometry with finite pitch and radius as a regular packing solution. The
resulting helix is energetically achiral, with left- and right-handed conformations exactly
degenerate in free energy; handedness is therefore selected spontaneously (spontaneous
chiral symmetry breaking) in the absence of any explicit microscopic chirality. The
conceptual novelty of Route~A is that it produces helix selection \emph{without}
introducing chemical periodicity or sticker-like interactions: the selection arises from
geometry, packing, and excluded-volume constraints alone.

(ii) In Route~B, helix stabilization arises from periodic attractive interactions along the
backbone. Monomers separated by a fixed contour distance experience an additional energetic
gain, favoring repeated contacts that are naturally satisfied by a helical geometry. In this
case, helix formation follows from a geometric commensurability condition between the
backbone separation and the helical repeat, leading to a well-defined pitch and radius even
in the absence of steric thickness effects. This route connects naturally to the physics of
sequence-encoded polymers and to classical cooperative pictures of helix formation, but is
formulated here in a continuum-geometry and free-energy language suitable for comparing
\emph{global} morphologies.

A central objective of this paper is to make these mechanisms explicit and analytically
transparent. We therefore develop the geometry of the helix, its curvature and torsion, and
the associated bending and cohesive contributions in a step-by-step manner. Particular
emphasis is placed on identifying the minimal ingredients required for helix stabilization
and on formulating simple scaling criteria for crossover against competing morphologies such
as rods and toroids, in terms of the persistence length, attraction strength, and chain
length. We also emphasize a diagnostic aspect: the same framework explains why many
standard models (including those studied in Ref.~\cite{SrinivasBagchi2002}) do \emph{not}
produce helices, while identifying the additional structure needed for helices to appear,
consistent with the ordered collapse observed in systems such as Ref.~\cite{HuBarbaraRosskyBagchi2000}.

\subsection*{Historical perspective: helix--coil theories of Zimm--Bragg and Gibbs--DiMarzio}

Before proceeding, it is useful to place the present work in the broader historical context
of helix--coil transitions in polymers. Two foundational theoretical frameworks deserve
special mention: the Zimm--Bragg theory and the Gibbs--DiMarzio approach
\cite{ZimmBragg1958,ZimmBragg1959,GibbsDiMarzio1958,GibbsDiMarzio1959}. 
Although developed originally with biopolymers in
mind, these theories introduced concepts that remain central to the statistical mechanics
of conformational ordering in one-dimensional systems, especially the ideas of
\emph{cooperativity} and of sharp but finite transitions in finite chains.

The Zimm--Bragg theory formulated the helix--coil transition as a one-dimensional
statistical-mechanical problem in which each repeat unit can exist in either a helical or a
coil state. The theory introduced two key parameters: a propagation parameter, which
controls the free-energy difference between helical and coil states, and a nucleation
parameter, which penalizes the initiation of a helical segment. This simple framework
captured the essential features of cooperativity, sharp transitions in finite chains, and
the emergence of long helical domains despite the absence of true phase transitions in one
dimension. Subsequent refinements, most notably the Lifson--Roig formulation
\cite{LifsonRoig1961}, clarified the role of nucleation and end effects and placed the theory
on a more systematic footing.

The Gibbs--DiMarzio approach, developed in parallel, emphasized a complementary physical
viewpoint. Rather than focusing primarily on transfer matrices, Gibbs and DiMarzio framed
helix formation as a balance between configurational entropy and energetic stabilization.
In this picture, helix formation represents a loss of conformational entropy that must be
compensated by favorable interactions, such as hydrogen bonding or other specific
intra-chain contacts. This perspective proved influential not only for helix--coil
transitions but also for later developments in polymer physics and glass theory, where
entropy--energy competition plays a central role.

It is important to emphasize that the present work addresses a fundamentally different, yet
conceptually related, problem. Classical Zimm--Bragg and Gibbs--DiMarzio theories focus on
\emph{local} conformational ordering along the polymer backbone, typically assuming that a
helical state is an internal degree of freedom available to each repeat unit. In contrast,
the helices studied in this paper are \emph{global condensate morphologies} of an entire
polymer chain. Here, helix formation is not an internal conformational choice made locally
along the backbone, but a three-dimensional packing geometry that must compete, at the
level of free energy, with other collapsed morphologies such as globules, toroids, and
rods.

Nevertheless, there is a deep conceptual connection between these classical theories and
the mechanisms explored in the present work. The ideas of cooperativity, of finite
nucleation barriers, and of sharp but finite transitions in finite systems reappear
naturally in our analysis, particularly in Route~(B), where periodic interactions along
the backbone stabilize helical order through collective effects. In this sense, the
present theory may be viewed as extending the spirit of helix--coil theories from local
ordering phenomena to the level of global condensed morphologies, while remaining firmly
grounded in the language of geometry and free energy.

%


\section{Helical geometry and energetic ingredients: Equation of helix}

Despite its exotic conformation, helix is described by a simple set of equations.

We describe a helical polymer conformation as a space curve
$\bm r(s)$ parameterized by arc length $s\in[0,L]$ with $L=Nb$.
A circular helix of radius $R$ and pitch $P$ (advance per $2\pi$ rotation)
may be written as
\begin{equation}
\bm r(s)=
\big(
R\cos(\Omega s),\;
R\sin(\Omega s),\;
\alpha s
\big),
\end{equation}
where $\Omega$ is the \emph{angular rate} (radians per unit arc length) describing how fast
the curve winds around the helix axis as $s$ increases, and $\alpha$ is the \emph{axial rate}
(advance along the helix axis per unit arc length).
The constants $\Omega$ and $\alpha$ are fixed by the inextensibility condition
$|\partial_s \bm r|=1$:
\begin{equation}
(R\Omega)^2+\alpha^2=1.
\end{equation}
The pitch follows as
\begin{equation}
P=\alpha\,\frac{2\pi}{\Omega}.
\end{equation}
It is convenient to introduce the dimensionless helix parameter
\begin{equation}
u\equiv \frac{P}{2\pi R},
\end{equation}
in terms of which
\begin{equation}
R\Omega=\frac{1}{\sqrt{1+u^2}},
\qquad
\alpha=\frac{u}{\sqrt{1+u^2}}.
\end{equation}

As mentioned above,  $\Omega$ is the \emph{angular rotation rate} of the helix measured in radians
per unit contour length (i.e.\ per unit arc length $s$), while $\alpha$ is the
\emph{axial advance rate} (the $z$-component of the tangent).  Thus, as one moves
along the chain by an amount $ds$, the azimuthal angle increases by $\Omega\,ds$
and the chain advances along the helix axis by $\alpha\,ds$.  The curve is written
in an arc-length parametrization, so $s$ measures the true contour length along
the backbone.  The inextensibility condition $|\partial_s\bm r|=1$ therefore
enforces that the tangent vector has unit magnitude, and it is this constraint
that ties together the geometric parameters $(R,\Omega,\alpha)$.

It is also useful to anticipate the physical meaning of the dimensionless ratio
$u=P/(2\pi R)$ introduced above.  Since $2\pi R$ is the circumference of one turn,
$u$ measures the pitch relative to the turn circumference.  Small $u$ corresponds
to a \emph{tight helix} (small pitch, large overlap of turns), while large $u$
corresponds to a \emph{stretched helix} approaching a nearly straight rod.
Many selection rules and crossover criteria are most transparent when expressed
in terms of $u$ rather than in terms of $(\Omega,\alpha)$ separately.

\subsection {Curvature, torsion, and bending energy}

A useful geometric property of a helix is that both its curvature and its torsion
are constant along the contour, i.e., they do not depend on the arc-length coordinate $s$.
Since later free-energy
comparisons hinge on the fact that the curvature penalty is \emph{uniform along the
entire contour}, it is pedagogically useful to record (at least briefly) why the
expressions below follow. We use standard differential geometry needed to describe the helical curve
\cite{DifferentialGeo,LandauElasticity,Love1944}

For an arc-length parametrized curve, the curvature is defined by
$\kappa(s)=|\partial_s^2\bm r(s)|$, i.e.\ by the magnitude of the rate of change of
the unit tangent.  Differentiating the helix parametrization twice therefore yields
a purely transverse vector of magnitude $R\Omega^2$, which is independent of $s$,
so $\kappa$ is constant.  

The curvature and torsion of a space curve are defined by
\cite{DifferentialGeo}
\begin{equation}
\kappa = \left|\frac{d^2 \mathbf r}{ds^2}\right|,
\qquad
\tau = \frac{(\mathbf r' \times \mathbf r'')\cdot \mathbf r'''}
      {|\mathbf r' \times \mathbf r''|^2},
\end{equation}
where primes denote derivatives with respect to arc length $s$.
For the helical parametrization of Eq.~(1) one obtains directly
\begin{equation}
\kappa = R\Omega^2, \qquad \tau = \alpha \Omega .
\end{equation}

Using the inextensibility constraint $(R\Omega)^2+\alpha^2=1$
and the definition $u=P/(2\pi R)=\alpha/(R\Omega)$,
these expressions reduce to
\begin{equation}
\kappa_h=\frac{1}{R(1+u^2)}, \qquad
\tau_h=\frac{u}{R(1+u^2)} .
\end{equation}

The torsion $\tau(s)$ measures the rate at which the
osculating plane rotates along the curve; it may be computed from the standard
identity
\begin{equation}
\tau(s)=
\frac{\big(\partial_s\bm r\times\partial_s^2\bm r\big)\cdot\partial_s^3\bm r}
{\left|\partial_s\bm r\times\partial_s^2\bm r\right|^2},
\end{equation}
which again becomes constant for the helical ansatz.  The resulting formulas are
quoted below, and we emphasize that the constancy of $\kappa_h$ and $\tau_h$ is a
special feature of the helix that simplifies the bending functional drastically.

As mentioned above, a helix has constant curvature and torsion,
\begin{equation}
\kappa_h=\frac{1}{R(1+u^2)},
\qquad
\tau_h=\frac{u}{R(1+u^2)}.
\end{equation}

Here $\kappa(s)$ denotes the local curvature of the space curve $\bm r(s)$,
defined (for arc-length parametrization) by
\begin{equation}
\kappa(s)\equiv \left|\partial_s^2 \bm r(s)\right|
= \left|\frac{d\bm t}{ds}\right|,
\qquad \bm t(s)\equiv \partial_s \bm r(s),\quad |\bm t|=1.
\end{equation}

For a wormlike chain with persistence length $L_p$ and bending modulus
$\kappa_b=k_BT L_p$, the bending energy is
\begin{equation}
F_{\rm bend}
=
\frac{\kappa_b}{2}\int_0^L ds\,\kappa^2(s)
=
\frac{k_BT L_p}{2}\,L\,\frac{1}{R^2(1+u^2)^2}.
\label{eq:bend_helix_compact}
\end{equation}

\subsection{Distance between two points along the helix}

Since we need to calculate the interaction energy between beads placed along
the helical curve, we require the spatial distance between two points whose
contour coordinates differ by $\Delta s$.

Many steric constraints and ``sticker'' contact conditions reduce to a geometric
statement about the separation of two points on the helix whose contour
positions differ by $\Delta s$. The derivation is elementary but worth keeping
explicit because the same formula plays two distinct roles in the paper:
(i) it controls nonlocal self-avoidance when the polymer is treated as a thick
tube, and (ii) it controls commensurate contacts when attractive ``stickers''
act between monomers separated by a fixed contour distance $m$.

Starting from $\Delta\bm r=\bm r(s+\Delta s)-\bm r(s)$, the transverse part contains
the trigonometric combination
\[
[\cos(\Omega(s+\Delta s))-\cos(\Omega s)]^2
+
[\sin(\Omega(s+\Delta s))-\sin(\Omega s)]^2
=2[1-\cos(\Omega\Delta s)] .
\]
The axial part contributes $(\alpha\Delta s)^2$. Substituting
$\Omega=1/[R\sqrt{1+u^2}]$ and $\alpha^2=u^2/(1+u^2)$ yields

\begin{equation}
d^2(\Delta s)
=
2R^2\!\left[
1-\cos\!\left(
\frac{\Delta s}{R\sqrt{1+u^2}}
\right)
\right]
+
\frac{u^2}{1+u^2}(\Delta s)^2.
\label{eq:helix_distance}
\end{equation}

This expression governs both steric self-avoidance and periodic-contact
conditions used later.

\subsection {Generic cohesive contributions}

At a coarse-grained level, attractive interactions favor local packing and may
be represented by an effective cohesive gain
\begin{equation}
F_{\rm coh}\sim -\varepsilon_{\rm eff}\,\mathcal N_c,
\end{equation}
where $\mathcal N_c$ is the number of effective near contacts.
In collapsed states one may also include an interfacial term $\gamma S$;
when comparing different ordered condensates, it is often convenient to
subtract a common dense bulk contribution.

At this stage it is important to stress a point that motivates the remainder of the
paper.  In a minimal collapsed-state free energy of the form
\emph{(surface tension)} $+\,$ \emph{(bending elasticity)}, there is no term that
rewards torsion, periodic registry, or regular packing beyond what is already
captured by generic cohesion and interfacial area.  A helix therefore pays the
bending penalty \emph{everywhere along the contour} (because $\kappa_h$ is constant),
but it does not gain any special energetic benefit for being twisted.
Competing morphologies can ``hide'' curvature: a rod suppresses curvature almost
entirely, while a toroid confines curvature to a well-defined radius and can
optimize it together with surface terms.  This is the fundamental reason helices
are generically absent in standard bead--spring collapse models with isotropic
attractions, and why an additional \emph{helix-selecting} mechanism is required.

The two mechanisms developed later may be viewed as minimal ways of supplying such
a selection principle without appealing to biochemical specificity.  Route (A) uses
a thickness/tube constraint so that sterics and packing \emph{geometrically} select a
finite pitch and radius; Route (B) uses periodic interactions so that \emph{energetic
commensurability} selects a helical repeat.  The present section provides the
geometric and energetic building blocks common to both routes.

\subsection{Cooperative stabilization and relation to classical helix--coil theories}
\label{subsec:cooperativity}

The periodic--sticker mechanism introduced above shares an important conceptual
feature with classical helix--coil theories, most notably those of Gibbs--DiMarzio
and Zimm--Bragg.  In those theories, helix stabilization arises from hydrogen
bonding between repeat units separated by a fixed contour distance along the
backbone (for example, $i$ and $i+3$ or $i+4$ in polypeptides).  Helicity is thus
encoded through \emph{regularly spaced internal bonds}, rather than through generic
isotropic attraction.

A key consequence of this construction is cooperativity.  An isolated hydrogen
bond does not define a stable helix; instead, several consecutive bonds (typically
three) must be formed to lock in helical order.  When such a contiguous set of bonds
is disrupted, a large configurational entropy is released.  This collective
entropic effect underlies the sharp but finite helix--coil transitions captured by
Gibbs--DiMarzio and Zimm--Bragg theories, and is often represented through a
nucleation penalty or cooperativity parameter.

The present periodic--sticker model incorporates the same physical idea, but at a
coarse-grained, field-theoretic level appropriate for global condensate
morphologies.  Stickers are placed at a fixed contour separation $m$, and helix
stabilization requires that the geometric commensurability condition be satisfied
over \emph{several consecutive sticker pairs}.  In this sense, only clusters of
satisfied stickers contribute coherently to helix stabilization, while isolated
satisfied contacts do not.  The free-energy gain associated with stickers therefore
reflects not only the energetic strength of individual interactions but also their
collective, geometry-enabled reinforcement along the chain.

From this viewpoint, the periodic--sticker mechanism may be regarded as a
continuum, geometric analogue of classical helix--coil cooperativity.  The
essential ingredients---fixed contour separation, multi-bond stabilization, and
large entropy release upon collective bond disruption---are preserved, while the
description remains compatible with the variational and field-theoretic framework
used to analyze competing condensed morphologies such as rods and toroids.
%

\begin{figure}[t]
\centering
\includegraphics[width=0.9\linewidth]{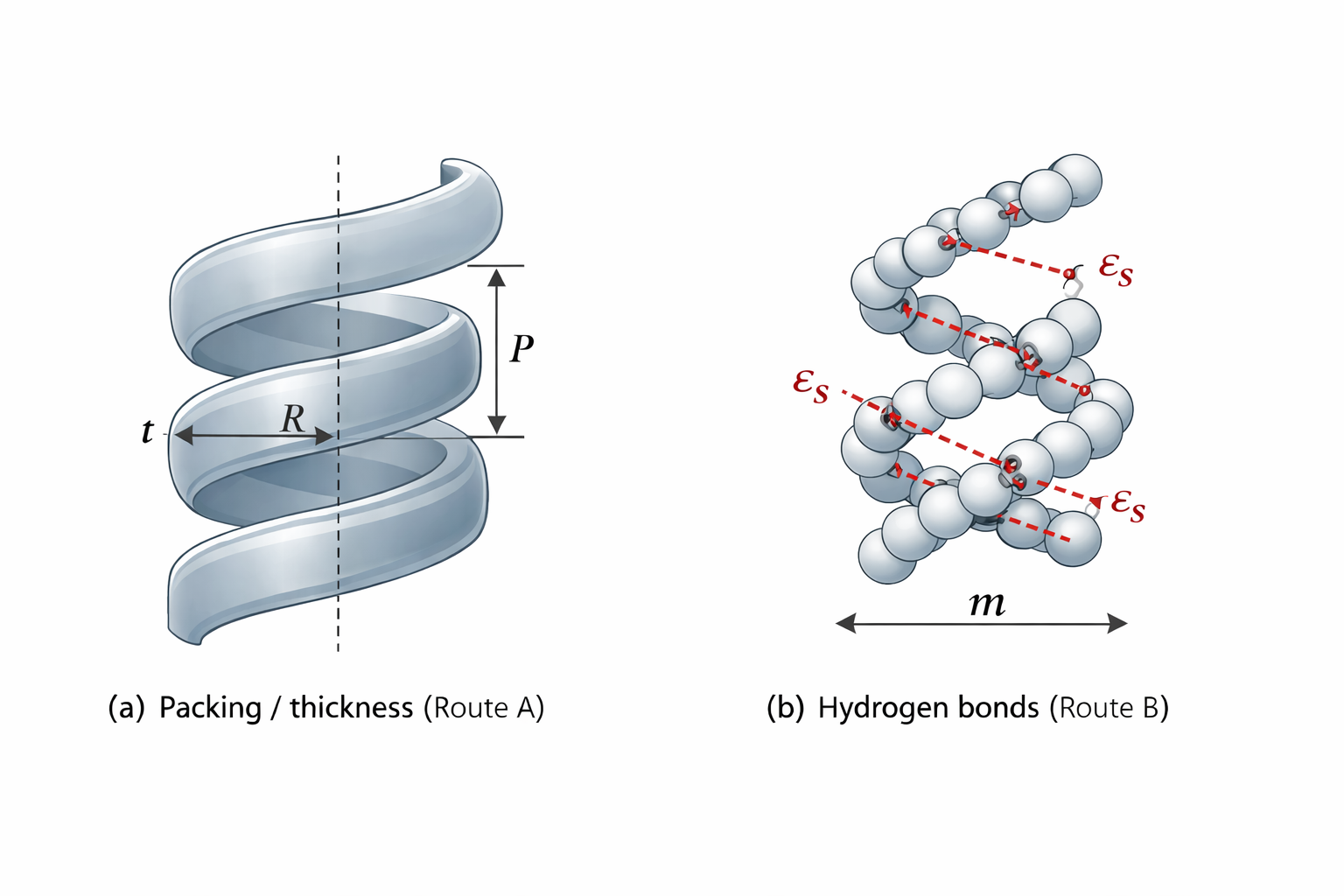}

\caption{
Two minimal routes to helix formation in collapsed polymers.
(a) Packing/thickness route (Route A): a polymer treated as a tube of
finite thickness $t$ forms a helix characterized by radius $R$ and pitch $P$
when packing and curvature constraints are satisfied.
(b) Hydrogen-bond / sticker route (Route B): periodic attractive
interactions between monomers separated by a contour distance $m$
stabilize a helical registry through repeated contacts of strength
$\varepsilon_s$.
}

\label{fig:helix_routes}
\end{figure}
%

\section{Periodicity-assisted helix stabilization: commensurability and stickers (Route B)}
\label{sec:routeB}
Having established that helices can arise purely from geometric and steric
constraints, we now examine how \emph{periodic internal interactions} further
stabilize and refine helical order.
This mechanism, Route~(B), is closely connected to classical helix--coil theories
(Gibbs--DiMarzio and Zimm--Bragg), but is formulated here within a continuum,
geometric framework appropriate for global polymer morphologies.

\subsection{Periodic sticker contact condition}

We consider attractive ``sticker'' interactions between monomers separated by a
fixed contour distance $m$ (in units of the step length $b$).
For two such monomers, separated by $\Delta s=mb$, the helix distance formula gives
\begin{equation}
d^2(mb)
=
2R^2\!\left[1-\cos\!\left(\frac{mb}{R\sqrt{1+u^2}}\right)\right]
+
\frac{u^2}{1+u^2}(mb)^2 .
\label{eq:dmbB}
\end{equation}
A sticker bond is considered satisfied when
\begin{equation}
d(mb)\le a_c ,
\label{eq:contactB}
\end{equation}
where $a_c$ is a capture distance of order the monomer size.
Equations~\eqref{eq:dmbB} and \eqref{eq:contactB} define the feasible region in
$(R,u)$ space for periodic contacts.

\subsection{Commensurability and axial separation}

A particularly transparent regime occurs when the angular advance over $mb$ is
close to an integer number of turns,
\begin{equation}
\frac{mb}{R\sqrt{1+u^2}} \approx 2\pi n ,
\qquad n\in\mathbb{Z}^{+}.
\label{eq:commB}
\end{equation}
Writing $2\pi n+\delta$ with $|\delta|\ll1$ and expanding gives
\begin{equation}
d^2(mb)\simeq R^2\delta^2 + \frac{u^2}{1+u^2}(mb)^2 .
\label{eq:dmb_commB}
\end{equation}
In the commensurate limit $\delta\simeq0$, the distance is dominated by axial
separation,
\begin{equation}
d(mb)\approx \frac{u}{\sqrt{1+u^2}}\,mb .
\label{eq:axialB}
\end{equation}
The sticker condition \eqref{eq:contactB} therefore implies
\begin{equation}
\frac{u}{\sqrt{1+u^2}} \lesssim \frac{a_c}{mb},
\label{eq:uineqB}
\end{equation}
which, for $a_c\ll mb$, requires small pitch ($u\ll1$).

\subsection{Dimensionless formulation}

For quantitative analysis, define
\begin{equation}
x\equiv \frac{mb}{R\sqrt{1+u^2}},
\qquad
R=\frac{mb}{x\sqrt{1+u^2}}.
\end{equation}
Equation~\eqref{eq:dmbB} then becomes
\begin{equation}
\frac{d^2(mb)}{(mb)^2}
=
\frac{2}{x^2(1+u^2)}(1-\cos x)
+
\frac{u^2}{1+u^2}.
\label{eq:dimlessB}
\end{equation}
Setting $d(mb)=a_c$ yields the master constraint
\begin{equation}
\frac{a_c^2}{(mb)^2}
=
\frac{2}{x^2(1+u^2)}(1-\cos x)
+
\frac{u^2}{1+u^2},
\label{eq:masterB}
\end{equation}
which implicitly determines the allowed helix parameters.

\subsection{Sticker contribution and stability}

If each satisfied sticker contributes an energy $-\varepsilon_s$, the maximum
number of non-overlapping periodic contacts is of order $N/m$, giving
\begin{equation}
F_{\rm sticker}\approx -\varepsilon_s\,\frac{N}{m}\,\mathcal{P}(R,u),
\label{eq:FstickerB}
\end{equation}
where $\mathcal{P}(R,u)$ is the fraction of satisfied contacts.
The helix free energy becomes
\begin{equation}
F_{\rm helix}^{(B)}
=
\frac{k_BT L_p}{2}\,L\,\frac{1}{R^2(1+u^2)^2}
+F_{\rm base}
-\varepsilon_s\,\frac{N}{m}\,\mathcal{P}(R,u).
\label{eq:FhelixB}
\end{equation}
If $\mathcal{P}\approx1$, comparing sticker gain with bending cost yields
\begin{equation}
\varepsilon_s
\gtrsim
\frac{k_BT L_p}{2}\,
\frac{mb}{R^2(1+u^2)^2}.
\label{eq:thresholdB}
\end{equation}
In the commensurate limit $x\approx 2\pi n$,
\begin{equation}
R \approx \frac{mb}{2\pi n\sqrt{1+u^2}},
\end{equation}
which gives, for $u\ll1$,
\begin{equation}
\varepsilon_s
\gtrsim
2\pi^2 n^2\,\frac{k_BT L_p}{m b}.
\label{eq:thresholdB_simple}
\end{equation}

\subsection{Cooperativity and relation to classical helix--coil theories}

Periodic stickers introduce cooperativity analogous to that in
Gibbs--DiMarzio and Zimm--Bragg theories.
Helix stabilization requires that the commensurability condition be satisfied
over several consecutive sticker pairs; isolated satisfied contacts do not
contribute coherently.
Breaking a contiguous cluster of such contacts releases a large configurational
entropy, leading to sharp but finite helix--coil transitions.
The present formulation captures this physics within a continuum geometric
framework, without invoking discrete transfer-matrix descriptions.


\section{Free-energy consolidation and morphological selection}
\label{sec:free_energy_selection}

Having developed the geometry of helices (Section~2), and established two distinct
helix-selecting mechanisms—Route~(A) based on packing/thickness constraints
(Section~2) and Route~(B) based on periodic interactions and commensurability
(Section~3)—we now consolidate the theory at the free-energy level.
The goal of this section is not to re-derive geometric results, but to place helices
on the same thermodynamic footing as rods and toroids and to formulate clear
criteria for morphological crossovers.

\subsection{Helices in coarse-grained theories}

In a minimalistic coarse-grained description including only bending elasticity and isotropic cohesion
(surface tension), collapsed polymers preferentially adopt morphologies that
minimize curvature penalties at fixed volume.
Rods and toroids naturally emerge in this framework: rods minimize curvature almost
entirely, while toroids localize curvature in a controlled manner.

Helices, by contrast, carry a \emph{uniform curvature} throughout their contour.
In the absence of an additional compensating mechanism, a helix therefore incurs a
bending penalty without receiving a corresponding energetic reward.
This observation explains why helices do not appear in minimal
``surface tension + bending'' theories and motivates the search for helix-specific
stabilization mechanisms.

\subsection{Helix free energies from Route (A) and Route (B)}

\paragraph{Route (A): packing/thickness-controlled helices.}

As we mentioned earlier, this aspect of helix formation has not found much
attention in the past.

When the polymer has a finite effective thickness $t$, steric constraints impose
both a local curvature bound and a nonlocal self-avoidance condition.
As shown in Section~2, the combined local curvature and nonlocal self-avoidance
constraints fix the tightly packed helix geometry to

\begin{equation}
R \sim t,
\qquad
P \sim 2t,
\qquad
u = \mathcal{O}(1).
\end{equation}
The corresponding helix free energy may be written, up to baseline dense-state
contributions common to all collapsed morphologies, as
\begin{equation}
F_{\rm helix}^{(A)}
\sim
\frac{k_B T L_p}{2}\,\frac{L}{t^2}
-
\varepsilon_{\rm eff}\,\frac{L}{b},
\label{eq:FhelixA_consolidated}
\end{equation}
where $L_p$ is the persistence length and $\varepsilon_{\rm eff}$ is an effective
cohesive energy per segment.
Comparison with a rodlike reference state (negligible bending cost) yields the
crossover condition
\begin{equation}
\varepsilon_{\rm eff}
\gtrsim
\frac{k_B T L_p}{2}\,\frac{b}{t^2}.
\label{eq:crossoverA}
\end{equation}
Thus, Route~(A) predicts a helix window controlled primarily by the ratio
$L_p/t^2$ and the strength of generic attractions.
Importantly, this mechanism is \emph{purely geometric} and does not rely on
chemical periodicity or sequence specificity.

\paragraph{Route (B): periodic sticker–stabilized helices.}

This is the more common scenario we encounter often in biomolecules.

In Route~(B), helices are stabilized by attractive interactions between monomers
separated by a fixed contour distance $m$.
When the helix geometry satisfies the commensurability condition
\begin{equation}
d(mb) \lesssim a_c,
\end{equation}
each satisfied sticker contributes an energy $-\varepsilon_s$.
The resulting energetic gain scales as
\begin{equation}
F_{\rm sticker}
\sim
-\varepsilon_s\,\frac{N}{m}.
\label{eq:Fsticker_consolidated}
\end{equation}
Balancing this gain against the bending cost of the helix yields, in the
commensurate and small-pitch limit, the stability criterion
\begin{equation}
\varepsilon_s
\gtrsim
2\pi^2\,\frac{k_B T L_p}{m b},
\label{eq:crossoverB}
\end{equation}
up to numerical factors of order unity.
This route therefore favors either strong periodic interactions or large sticker
spacing $m$ and is most naturally realized in sequence-encoded polymers.
Unlike Route~(A), the stabilization mechanism here is explicitly energetic and
commensurate rather than geometric.

\subsection{Embedding helices into the full morphology diagram}

To determine when helices become stable relative to rods, toroids, or the
extended coil, it is necessary to place all candidate structures within a
common thermodynamic framework.  In particular, the comparison must be made
relative to a well-defined \emph{reference state} of the collapsed polymer.

In the poor–solvent regime the polymer interior is well approximated by a
uniform dense phase whose free-energy density depends only weakly on the
global shape of the chain.  Different morphologies therefore share a common
bulk contribution proportional to the polymer volume (or equivalently the
number of monomers).  Since this term is identical for all collapsed states,
it does not influence which morphology is selected.  It is therefore convenient
to subtract this common bulk term and work with \emph{reduced free energies}
that contain only the geometry-dependent contributions.  These include surface
terms, bending penalties, and additional interaction energies specific to a
given morphology.  This procedure places rods, toroids, and helices on equal
thermodynamic footing and isolates the energetic ingredients that actually
control morphological selection.

Within this baseline description the construction of a morphology diagram
proceeds in several steps.

\paragraph{Step 1: baseline model.}
Adopt the same dense-phase background used for globules, toroids, and rods.
The polymer interior is treated as a uniform condensed phase with an effective
surface tension $\gamma$ associated with polymer–solvent interfaces.  A common
bulk contribution is subtracted so that only geometry-dependent terms remain
in the comparison between morphologies.

\paragraph{Step 2: candidate free energies.}
Construct reduced free energies for each possible morphology:
\begin{itemize}
\item Rod: $F'_{\rm rod}$ determined primarily by surface contributions from the
cylindrical body and end caps, with optional corrections from orientational
ordering when the chain becomes locally aligned.

\item Toroid: $F'_{\rm tor}$ containing bending energy of the curved chain
together with the surface energy of the toroidal condensate, minimized with
respect to the major radius.

\item Helix (Route A): $F^{(A)}_{\rm helix}$ given by
Eq.~\eqref{eq:FhelixA_consolidated}.  In this case the helix radius and pitch
are essentially fixed by steric thickness constraints, so the geometry is
largely predetermined by packing.

\item Helix (Route B): $F^{(B)}_{\rm helix}(R,u)$ derived in Section~4.  Here the
geometry is determined by minimizing the free energy subject to the
commensurate sticker–contact condition imposed by the interaction spacing $m$.
\end{itemize}

\paragraph{Step 3: minimization.}
Each candidate free energy is minimized with respect to its remaining geometric
degrees of freedom.  For rods this involves optimizing the aspect ratio, for
toroids the major radius, and for Route~(B) helices both the radius and the
pitch parameter $u$, subject to the appropriate constraints.

\paragraph{Step 4: defining crossovers.}
Morphological transitions are determined by equality of the minimized free
energies.  In practice the relevant crossover conditions are
\begin{equation}
F'_{\rm helix} = F'_{\rm tor},
\qquad
F'_{\rm helix} = F'_{\rm rod},
\qquad
F_{\rm coil} = F_{\rm helix},
\end{equation}
where the last equality becomes relevant when the extended coil competes
directly with a helical condensate.

Within this framework the two helix-forming mechanisms lead to distinct
stability criteria.  Route~(A) predicts helix formation primarily as a
function of the geometric stiffness ratio $L_p/t^2$ together with the overall
cohesive strength of the condensed phase.  Route~(B), by contrast, predicts
helix stability controlled by the sticker interaction strength
$\varepsilon_s$, the contour separation $m$, and the geometric
commensurability between the helical repeat and the interaction spacing.

\subsection{Physical interpretation}

Helices are distinguished from rods and toroids by the fact that they are never
selected by surface tension and bending alone.
They become stable only when an additional helix-specific mechanism is operative.
In Route~(A) this mechanism is geometric and steric, arising from thickness and
packing constraints.
In Route~(B) it is energetic and periodic, arising from commensurate interactions
along the backbone.
The competition and possible coexistence of these mechanisms determine the
location and extent of helical regions within the overall polymer morphology
diagram.
 \section{Numerical estimates and quantitative predictions}
\label{sec:numerics_predictions}

The theory developed in Sections~\ref{sec:routeA}--\ref{sec:routeB} yields explicit
free-energy inequalities for helix stabilization.  This enables quantitative
predictions and, importantly, a diagnostic interpretation of ``negative results'':
for example, the absence of helices in simulations that nevertheless show globules,
toroids, and rods.  Conversely, the same estimates clarify why helices are readily
observed in DNA-like systems, where periodic hydrogen bonding and cooperativity
produce strong, sequence-encoded stabilization.

Throughout this section we express energetic parameters in dimensionless form:
\begin{equation}
\ell_p \equiv \frac{L_p}{b},\qquad
\tilde t \equiv \frac{t}{b},\qquad
u_{\rm eff}\equiv \frac{\varepsilon_{\rm eff}}{k_B T},\qquad
u_s\equiv \frac{\varepsilon_s}{k_B T},
\end{equation}
where $L_p$ is the persistence length, $b$ is the segment length, $t$ is the
effective tube thickness, $\varepsilon_{\rm eff}$ is the effective cohesive gain
(per segment) in Route~(A), and $\varepsilon_s$ is the sticker strength in Route~(B).

\subsection{Route (A): geometric helix window controlled by $\ell_p/\tilde t^2$}

Route~(A) predicts that helix stabilization relative to a low-curvature reference
state (rodlike, or a toroid in the appropriate limit) requires
\begin{equation}
u_{\rm eff}\ \gtrsim\ u_{\rm eff}^{\rm(th)}
\equiv \frac{\ell_p}{2\,\tilde t^{\,2}}.
\label{eq:ueff_threshold_numeric}
\end{equation}
This criterion makes immediate physical sense: increasing stiffness (larger $\ell_p$)
raises the bending penalty, whereas increasing thickness (larger $\tilde t$) reduces
the curvature cost by selecting a larger helix radius.

Table~\ref{tab:RouteA_thresholds} lists $u_{\rm eff}^{\rm(th)}$ for representative
values $\ell_p=4$--$7$ (a range relevant to several coarse-grained semiflexible
polymer models) and for thickness values $\tilde t=0.8$--$1.5$.  Even modest
changes in $\tilde t$ have a large effect because the threshold scales as $1/\tilde t^2$.

\begin{table}[t]
\centering
\caption{Route (A) helix-stability threshold
$u_{\rm eff}^{\rm(th)}=\ell_p/(2\tilde t^2)$ for representative stiffness
$\ell_p=L_p/b$ and effective thickness $\tilde t=t/b$.  Helices from Route~(A)
become plausible when the effective cohesive gain $u_{\rm eff}=\varepsilon_{\rm eff}/k_BT$
exceeds these values.}
\label{tab:RouteA_thresholds}
\vspace{0.15cm}
\begin{tabular}{c|cccc}
\hline
$\ell_p$ & $\tilde t=0.8$ & $\tilde t=1.0$ & $\tilde t=1.2$ & $\tilde t=1.5$ \\
\hline
4 & 3.13 & 2.00 & 1.39 & 0.89 \\
5 & 3.91 & 2.50 & 1.74 & 1.11 \\
6 & 4.69 & 3.00 & 2.08 & 1.33 \\
7 & 5.47 & 3.50 & 2.43 & 1.56 \\
\hline
\end{tabular}
\end{table}

\paragraph{Interpretation and connection to simulations without helices.}
In many bead--spring simulations (including those that robustly show globules,
toroids, and rods), generic attraction strengths are often parameterized through
Lennard--Jones-like well depths.  While the bare LJ depth can be several $k_BT$,
the quantity relevant to Eq.~\eqref{eq:ueff_threshold_numeric} is the \emph{effective}
cohesive gain per segment that is available \emph{after} accounting for excluded volume,
packing frustration, and the fact that competing morphologies (toroid/rod) already
achieve extensive contacts with less curvature cost.

Thus, even if the microscopic attraction scale is ``large,'' the \emph{morphology-selecting}
incremental gain that must compensate uniform helix curvature can be smaller.
Table~\ref{tab:RouteA_thresholds} shows that for moderately stiff chains
($\ell_p\simeq 4$--$7$) and modest thickness ($\tilde t\lesssim 1$),
the required $u_{\rm eff}^{\rm(th)}$ is of order $2$--$5$.
If the effective gain available uniquely to the helix falls below this window,
the system will naturally realize globules/toroids/rods but not helices---precisely
the pattern reported in several studies of semiflexible collapse.
In this sense, the absence of helices is not a failure of the model; it is a
quantitative signature that the system resides outside the Route~(A) helix window.

\subsection{Route (B): sticker window controlled by $\ell_p/m$ and commensurability}

For Route~(B), in the commensurate and small-pitch regime, the stability estimate reads
\begin{equation}
u_s\ \gtrsim\ u_s^{\rm(th)} \equiv 2\pi^2\,\frac{\ell_p}{m}
\;\;\approx\;\; 19.74\,\frac{\ell_p}{m}.
\label{eq:us_threshold_numeric}
\end{equation}
The key prediction is the strong $1/m$ dependence: periodic stabilization becomes
rapidly easier as the sticker spacing increases.

Table~\ref{tab:RouteB_thresholds} lists $u_s^{\rm(th)}$ for $\ell_p=4$--$7$ and
representative sticker spacings $m=10$--$40$.

\begin{table}[t]
\centering
\caption{Route (B) helix-stability threshold
$u_s^{\rm(th)}\approx 2\pi^2\,\ell_p/m$ for representative stiffness $\ell_p=L_p/b$
and sticker spacing $m$.  Helices from Route~(B) become plausible when sticker strength
$u_s=\varepsilon_s/k_BT$ exceeds these values \emph{and} the commensurability constraint
is satisfiable for some $(R,u)$.}
\label{tab:RouteB_thresholds}
\vspace{0.15cm}
\begin{tabular}{c|cccc}
\hline
$\ell_p$ & $m=10$ & $m=20$ & $m=30$ & $m=40$ \\
\hline
4 & 7.90 & 3.95 & 2.63 & 1.97 \\
5 & 9.87 & 4.94 & 3.29 & 2.47 \\
6 & 11.84 & 5.92 & 3.95 & 2.96 \\
7 & 13.82 & 6.91 & 4.61 & 3.45 \\
\hline
\end{tabular}
\end{table}

\paragraph{Why DNA-like systems show helices.}
In DNA-like or hydrogen-bond-dominated systems, sticker energies can be several
$k_BT$ or more (and can be effectively stronger when multiple bonds are jointly
stabilized).  Moreover, the underlying interaction pattern is inherently periodic,
so that the commensurability requirement is physically meaningful rather than accidental.
Table~\ref{tab:RouteB_thresholds} shows that for $m\gtrsim 20$, the required
$u_s^{\rm(th)}$ lies in the range $\sim 4$--$7$ for $\ell_p=4$--$7$, which is
consistent with strong, specific interactions such as hydrogen bonding.

\paragraph{Cooperativity (GD/ZB) and an effective lowering of the threshold.}
A central feature of Gibbs--DiMarzio and Zimm--Bragg theories is that helix formation
is cooperative: stable helical order requires several consecutive bonds, and disruption
releases entropy in a collective manner.  In the present continuum description, this
can be represented by recognizing that the \emph{effective} stabilization is not that
of a single sticker in isolation, but of a \emph{run} of $q$ consecutive satisfied
contacts.  At the free-energy level, this can be encoded through an effective
cooperativity factor $\mathcal{C}_q$ multiplying the sticker gain,
\begin{equation}
F_{\rm sticker}\sim -\mathcal{C}_q\,\varepsilon_s\,\frac{N}{m},
\qquad \mathcal{C}_q \ge 1,
\end{equation}
where $\mathcal{C}_q$ increases when consecutive contacts reinforce one another
(geometrically and entropically).  In such a cooperative regime, the practical
sticker threshold is reduced relative to Eq.~\eqref{eq:us_threshold_numeric} by
a factor of order $\mathcal{C}_q$:
\begin{equation}
u_s^{\rm(th,\,coop)} \sim \frac{1}{\mathcal{C}_q}\,u_s^{\rm(th)}.
\end{equation}
This explains why helical order can be robust in DNA-like systems even when a
single-bond estimate appears marginal: cooperativity effectively amplifies the
periodic stabilization.

\subsection{A diagnostic “phase logic’’ and testable predictions}

The inequalities derived above provide more than simple stability criteria;
they suggest a practical \emph{diagnostic logic} for identifying which physical
mechanism is responsible for helix formation in a given system.  Because the
two routes depend on different geometric and energetic parameters, the
appearance (or absence) of helices relative to other collapsed morphologies
offers a clear signature of the underlying mechanism.

First, consider the common situation in which collapse produces globules,
toroids, or rods but \emph{no helices}.  In the present framework this outcome
has a simple interpretation: neither of the helix-stabilizing conditions is
satisfied.  In other words, the effective geometric parameter remains below
the threshold required for Route~(A),
$u_{\rm eff}<u_{\rm eff}^{(\mathrm{th})}$,
and/or the sticker interaction parameter remains below the threshold for
Route~(B),
$u_s<u_s^{(\mathrm{th})}$,
or the geometric commensurability condition cannot be satisfied.  This regime
is therefore expected for many generic semiflexible collapse models in which
only isotropic attraction and bending elasticity are present.

Second, if helices appear in the absence of any obvious periodic bonding
mechanism, the most natural interpretation is that the stabilization arises
through Route~(A), namely geometric packing controlled by the effective
polymer thickness.  In this case the theory predicts a strong dependence on the
excluded-volume diameter $t$.  Since the key control parameter scales as
$L_p/t^2$, even modest changes in the effective thickness---for example by
altering the bead diameter in a coarse-grained simulation---should shift the
stability window of helices significantly.

Third, if helices appear in polymers with sequence-encoded or periodic
interactions, this points instead to Route~(B).  In this mechanism the stability
of helices is governed by the strength of sticker interactions and by the
contour spacing $m$ between interacting sites.  The theory therefore predicts a
strong dependence of helix formation on $m$, with a threshold scaling
approximately as $1/m$, as well as a pronounced sensitivity to cooperativity,
which effectively amplifies the energetic stabilization of contiguous
sticker contacts.

These predictions can be tested directly in simulation.  For Route~(A), one may
scan the parameter space $(L_p,\tilde t)$ and determine whether helices appear
when the geometric threshold identified in Table~1 is crossed.  For Route~(B),
the relevant control variables are $(L_p,m,u_s)$, and the predicted morphology
boundaries can be compared with the thresholds summarized in Table~2.  In both
cases the key diagnostic is whether the observed transitions occur near the
theoretical stability limits derived above.


\section{Emergence of Chirality (handedness) in Gibbs--DiMarzio theory of helix formation}
\label{sec:GD_chirality}

In Route~(B), helix formation is driven by periodic, sequence-encoded
interactions along the polymer backbone, analogous to the hydrogen-bond
stabilization mechanism underlying the classical helix--coil theories of
Gibbs and DiMarzio (GD) and Zimm and Bragg (ZB).
In contrast to Route~(A), where the helix geometry emerges primarily from
packing and thickness constraints, Route~(B) introduces a preferred repeat
distance along the chain that favors the formation of a helical registry
between monomers separated by a fixed contour distance.

An important additional question then arises: how is the \emph{handedness}
(chirality) of the helix determined?  In the absence of microscopic chiral
interactions, left-handed and right-handed helices are energetically
degenerate.  The theory developed here therefore does not attempt to explain
the microscopic origin of molecular chirality.  Instead, it addresses a
different but complementary question: once a helical state forms, how is a
single handedness stabilized and propagated along the chain?  As we show
below, the cooperative physics underlying the GD helix--coil transition
naturally suppresses handedness reversals and therefore produces long
domains of uniform chirality.

\subsection{Local states and cooperative units}

We describe the polymer at the level of coarse-grained backbone units
(monomers) indexed by $i=1,\ldots,N$.  Each unit may exist in one of three
local states
\begin{equation}
\sigma_i \in \{ C,\; L,\; R \},
\end{equation}
where $C$ denotes a coil-like (non-helical) configuration, while $L$ and
$R$ denote participation in a left-handed or right-handed helical registry,
respectively.

Helical order is stabilized by periodic interactions between monomers
separated by $m$ units along the chain contour, consistent with the GD
picture of hydrogen bonding between residues $i$ and $i+m$.  Formation of
a stable helical segment requires establishing a \emph{cooperative unit}
involving several consecutive residues.  Following the classical GD
framework, we denote the free-energy cost of nucleating such a unit by
$\Delta g_n$ and define the nucleation parameter
\begin{equation}
\sigma \equiv e^{-\beta \Delta g_n},
\qquad 0 < \sigma \ll 1 .
\end{equation}
The small value of $\sigma$ reflects the fact that initiating a helical
segment requires overcoming an entropic penalty associated with restricting
backbone configurations.

Once a helical segment is established, propagation of the helix along the
chain becomes energetically favorable.  The free-energy change associated
with adding one residue to an existing helical segment is denoted
$\Delta g_p$, with Boltzmann weight
\begin{equation}
s \equiv e^{-\beta \Delta g_p}.
\end{equation}
Helical order is favored when $s>1$, and the small nucleation parameter
$\sigma$ ensures the cooperative nature of helix formation: helices tend
to appear as contiguous segments rather than isolated residues.

\subsection{Handedness reversals and domain-wall free energy}

A crucial new element of the present formulation is the explicit treatment
of \emph{helical handedness}.  In the absence of microscopic chiral
interactions, the free energies of left- and right-handed helices are
identical,
\begin{equation}
F_L = F_R .
\end{equation}
Helix formation therefore involves a form of spontaneous symmetry breaking:
the first nucleated helical segment may adopt either handedness with equal
probability.  Once a handedness is selected locally, cooperative propagation
tends to extend that same handedness along the chain.

A change in chirality along the polymer cannot occur continuously.
Instead, a reversal $L \leftrightarrow R$ requires disrupting the local
helical registry over a finite segment of the chain.  Such a disruption
creates a \emph{domain wall} between two helical regions of opposite
handedness and therefore carries a free-energy cost.

Within the GD framework it is natural to identify this cost with the
free energy required to melt a cooperative unit of the helix.  We therefore
associate a handedness-reversal penalty
\begin{equation}
\Delta g_{LR} \simeq \Delta g_n ,
\label{eq:DG_LR_GD}
\end{equation}
since both processes involve breaking the same set of stabilizing
interactions and restoring configurational entropy.  The corresponding
Boltzmann weight
\begin{equation}
\omega \equiv e^{-\beta \Delta g_{LR}} \sim \sigma
\end{equation}
controls the probability of chirality reversals along the chain.

This result highlights an important physical connection: the same
cooperative interactions that sharpen the helix--coil transition also
suppress reversals of handedness.  When the nucleation penalty is large,
domain walls between $L$ and $R$ helices become rare, and long
single-handed helical domains emerge.

\subsection{Effect of a small chiral bias}

The analysis above assumes that left- and right-handed helices are exactly
degenerate.  In many real polymers, however, microscopic chirality of the
monomer units introduces a small energetic bias favoring one handedness
over the other.  For example, biological polymers built from L-amino acids
typically favor right-handed helices.

To represent such effects at a coarse-grained level, we introduce a small
bias $\Delta g_{\rm chiral}$ such that
\begin{equation}
F_L - F_R = \Delta g_{\rm chiral}.
\end{equation}
This bias is not derived here but is treated as an external parameter
reflecting the underlying molecular asymmetry of the monomers.

Even when $|\Delta g_{\rm chiral}|$ is small compared with $k_B T$, the
cooperative nature of helix formation strongly amplifies its effect.
Because domain walls between opposite-handed helices are costly, the
preferred handedness selected at nucleation can propagate over long
segments of the chain.  In this way a weak microscopic bias can produce a
macroscopically uniform helical chirality.

The present formulation therefore separates two distinct physical issues:
the microscopic origin of molecular chirality, which lies outside the
scope of the present theory, and the cooperative statistical-mechanical
mechanism by which a chosen handedness is stabilized and propagated along
a polymer chain.

\subsection{Transfer-matrix formulation}

The statistical mechanics of the chain can be formulated using a
nearest-neighbor transfer matrix acting on the local state space
$\{C,L,R\}$ introduced above.  A minimal matrix that incorporates
helix nucleation, propagation, and handedness reversals is

\begin{equation}
\mathbf T =
\begin{pmatrix}
1 & \sqrt{\sigma}\, s & \sqrt{\sigma}\, s \\
\sqrt{\sigma} & s & \omega s \\
\sqrt{\sigma} & \omega s & s
\end{pmatrix},
\label{eq:GD_transfer}
\end{equation}

where the basis ordering is $(C,L,R)$.  The parameter $\sigma$ controls
nucleation of helical segments, $s$ controls propagation of an existing
helix, and $\omega = e^{-\beta \Delta g_{LR}}$ represents the Boltzmann
weight associated with a reversal of handedness.

For a chain of $N$ units the partition function is

\begin{equation}
Z = \bm v^{\mathsf T} \mathbf T^{\,N} \bm u ,
\end{equation}

where $\bm v$ and $\bm u$ specify boundary conditions.  In the long-chain
limit the thermodynamics is governed by the largest eigenvalue
$\lambda_{\max}$ of $\mathbf T$,

\begin{equation}
f = -k_B T \ln \lambda_{\max},
\end{equation}

which gives the free energy per residue.

The average helical fraction is obtained from

\begin{equation}
\theta
= \frac{\langle N_L + N_R \rangle}{N}
= \frac{\partial \ln \lambda_{\max}}{\partial \ln s},
\end{equation}

while the characteristic handedness domain length follows from the
probability of chirality reversals,

\begin{equation}
\xi \sim \omega^{-1}
      \sim e^{\beta \Delta g_{LR}} .
\end{equation}

Thus strong cooperativity (large $\Delta g_{LR}$) leads to long
domains of uniform helical handedness.

\subsection{Effect of a weak chiral bias}

As discussed in the previous subsection, real polymers may possess a
small microscopic chiral bias favoring one handedness.  At the level of
the transfer-matrix description this bias can be represented by slightly
different propagation weights for the two helical states,

\begin{equation}
s_L = s\, e^{\beta h},
\qquad
s_R = s\, e^{-\beta h},
\end{equation}

where $h$ is a small chiral field.

The resulting handedness order parameter

\begin{equation}
\chi
= \frac{\langle N_L - N_R \rangle}{\langle N_L + N_R \rangle}
\end{equation}

is obtained from

\begin{equation}
\chi =
\frac{1}{\theta}
\frac{\partial \ln \lambda_{\max}}{\partial (\beta h)} .
\end{equation}

Although $h$ may be very small compared with $k_B T$, cooperativity
strongly amplifies its effect.  Over a domain of length $\xi$ the bias
produces an excess free energy of order $h\xi$, so that when
$h\xi \gg k_B T$ one handedness dominates over experimentally
relevant chain lengths even if the microscopic chiral bias is weak.

\subsection{Physical interpretation and predictions}

This GD-based formulation leads to several concrete predictions.  First, the
free-energy cost of helix reversals is directly tied to the cooperative
nucleation free energy, $\Delta g_{LR} \simeq \Delta g_n$.  Second, the typical
handedness domain size grows exponentially with this cooperative free energy,
$\xi \sim e^{\beta \Delta g_n}$.  Third, weak microscopic chirality is strongly
amplified by cooperativity, yielding robust single-handed helices in finite
chains.  These results provide a semi-microscopic foundation for chiral
symmetry breaking within Route (B), grounded in the classical Gibbs--DiMarzio
picture yet extending it to address handedness selection explicitly.
%
%

\section{Numerical estimates and experimental realization of helical chirality}
\label{sec:numerics_expt}

Helical chirality in polymers is often discussed at a qualitative or phenomenological
level, particularly in biopolymers where molecular chirality is built in.
One of the advantages of the present formulation is that it allows
\emph{quantitative estimates} of helical handedness selection and domain sizes
using experimentally accessible parameters.
In this section we provide numerical estimates based on realistic polymer
parameters, compare the predictions of Routes (A) and (B), and discuss
experimental systems where the predicted chiral amplification mechanism
should be observable.

\subsection{Parameter estimates and physical scales}

We begin by summarizing representative parameter values relevant to
synthetic and biological polymers.

\paragraph{Persistence length.}
For semiflexible polymers of interest,
\begin{equation}
L_p \simeq 4\!-\!7\,b
\end{equation}
is typical for flexible synthetic polymers and polypeptides,
while DNA-like systems have much larger stiffness,
$L_p/b \sim 50$--$150$ depending on ionic conditions.

\paragraph{Energetic scales.}
Generic non-specific attractions (e.g.\ Lennard--Jones or hydrophobic)
typically correspond to
\begin{equation}
\varepsilon_{\rm eff} \sim 1\!-\!3\,k_BT,
\end{equation}
as inferred from simulations and experiments on polymer collapse.
Hydrogen-bond or sticker-like interactions are substantially stronger,
\begin{equation}
\varepsilon_s \sim 3\!-\!10\,k_BT,
\end{equation}
consistent with estimates from peptide hydrogen bonding and
DNA base pairing.

\paragraph{Cooperativity parameters.}
In the Gibbs--DiMarzio framework, cooperativity is encoded in the
nucleation parameter
\begin{equation}
\sigma = e^{-\beta \Delta g_n},
\end{equation}
with $\sigma \sim 10^{-2}$--$10^{-4}$ for strongly cooperative helices
(polypeptides, DNA), and $\sigma \sim 10^{-1}$ for weakly cooperative
synthetic systems.

These numbers allow direct numerical evaluation of chiral domain sizes
and handedness amplification.

\subsection{Numerical estimates for chiral domain sizes}

Within the extended Gibbs--DiMarzio description developed in
Section~\ref{sec:GD_chirality}, the typical handedness domain length is
\begin{equation}
\xi \sim \omega^{-1} \sim e^{\beta \Delta g_{LR}} \sim e^{\beta \Delta g_n}.
\end{equation}

Using representative values:

\begin{itemize}
\item For $\Delta g_n \simeq 2\,k_BT$ (weak cooperativity):
\[
\xi \sim e^{2} \approx 7,
\]
so frequent left--right reversals occur and macroscopic chirality is weak.

\item For $\Delta g_n \simeq 4\,k_BT$:
\[
\xi \sim e^{4} \approx 55,
\]
leading to long helical segments of fixed handedness.

\item For $\Delta g_n \simeq 6\,k_BT$ (DNA-like cooperativity):
\[
\xi \sim e^{6} \approx 400,
\]
implying essentially single-handed helices over experimentally relevant
chain lengths.
\end{itemize}

These values demonstrate that \emph{cooperativity alone} can generate
macroscopic chirality, even when the microscopic chiral bias is extremely weak.

\subsection{Amplification of weak chiral bias}

A central prediction of the present theory is the exponential amplification
of microscopic chirality.
If a small chiral field $h$ favors one handedness, the excess free energy
over a domain of size $\xi$ scales as
\begin{equation}
\Delta F_{\rm chiral} \sim h\,\xi.
\end{equation}
Even for
\begin{equation}
h \sim 10^{-2}\,k_BT,
\end{equation}
which is typical of weak stereochemical biases,
a cooperative domain with $\xi \sim 100$ yields
\[
\Delta F_{\rm chiral} \sim k_BT,
\]
sufficient to suppress the unfavored handedness.
Thus, the observed chirality of helices need not reflect a strong microscopic
bias; instead, it emerges from the interplay of weak bias and strong cooperativity.

This mechanism provides a natural explanation for why biopolymers composed
of only weakly chiral monomers nonetheless form robustly single-handed helices.

\subsection{Comparison with simulations and experiments}

Early simulations by Srinivas and Bagchi demonstrated that semiflexible
polymers with isotropic attractions form globules, toroids, and rods,
but \emph{not helices}, consistent with the absence of Route~(B)-type
periodic interactions.
Helices emerge in simulations only when either tube-like thickness
constraints (Route~A) or explicit directional interactions are included.

Experimentally, DNA and polypeptides represent canonical realizations of
Route~(B).  Hydrogen bonding between residues separated by fixed contour
distances provides both periodicity and strong cooperativity, leading to
large $\xi$ and pronounced chiral amplification.
Synthetic analogues with programmable stickers or patchy interactions
offer promising platforms to test the present predictions quantitatively.

Observable signatures include:
\begin{itemize}
\item circular dichroism signals scaling with $\xi$,
\item suppression of handedness reversals with increasing cooperativity,
\item tunable chirality upon systematic variation of sticker strength or spacing.
\end{itemize}

\subsection{Summary of numerical predictions}

The numerical estimates presented here support three robust conclusions:
(i) helices stabilized solely by generic attractions are unlikely;
(ii) cooperative periodic interactions naturally generate large chiral
domains; and
(iii) weak microscopic chirality is exponentially amplified by cooperativity.
Together, these results elevate Route~(B) from a qualitative mechanism to a
quantitatively predictive theory of helical chirality in polymers.

%
%

\section{Conclusions}
\label{sec:conclusion}

Let us now summarize the main results obtained in this study.

In this work we have developed a unified theoretical framework for the formation of
helical polymer condensates and for their competition with more familiar collapsed
morphologies such as globules, toroids, and rods.
A central message that emerges from our analysis is that \emph{helix formation is not
a generic consequence of monomer--monomer attraction combined with chain bending
elasticity}.
While these minimal ingredients are sufficient to produce globules, toroids, and
rods, they do not, by themselves, stabilize helices.

This observation is physically transparent.
Globules, toroids, and rods are all morphologies that either avoid curvature entirely
(rod) or confine it to limited regions (toroid), thereby minimizing bending penalties
while exploiting isotropic cohesion.
A helix, by contrast, carries a \emph{uniform curvature along its entire contour}.
In a minimal ``surface tension plus bending'' description, this curvature cost is not
compensated by any special energetic gain.
As a result, helices are generically disfavored in free-energy competition.
The frequent absence of helices in simulations of semiflexible polymer collapse is
therefore not anomalous; it is the natural outcome of the minimal theory.
Recognizing this negative result is an essential starting point of the present work.

At the same time, helices are ubiquitous in specific physical and biological systems.
This apparent contradiction motivates the central contribution of this paper: the
identification and systematic analysis of two distinct physical routes by which
helices can become stable once the minimal description is augmented in well-defined
ways.

The first route, Route~(A), is geometric and steric in origin and represents a genuinely
new element in the theory of polymer condensation.
When the polymer backbone has a finite effective thickness, local curvature bounds and
nonlocal self-avoidance constraints act together to restrict the space of allowed dense
packings.
Remarkably, these purely geometric constraints select a helical conformation with
finite radius and pitch as an optimal packing solution.
Tight packing fixes these geometric parameters to scale with the thickness, leading to
a well-defined bending cost and an extensive cohesive gain.
No chemical periodicity, sequence specificity, or microscopic chirality is required.
Helices emerge here as a consequence of differential geometry and steric packing alone.

A key and perhaps counterintuitive result of Route~(A) is that helix formation precedes
chirality.
Left- and right-handed helices are exactly degenerate in free energy, and chirality
emerges only through spontaneous symmetry breaking once a helical geometry is selected.
This establishes that chiral order can arise in polymers even in the complete absence
of chiral interactions, purely as a consequence of geometric degeneracy and cooperative
stabilization.
Because it relies only on thickness and self-avoidance, Route~(A) is universal in
character and applies to generic thick polymers, providing a natural explanation for
helices observed in tube-like or sterically constrained chains.

The second route, Route~(B), is energetic and commensurate in nature.
Here helix stabilization arises from periodic attractive interactions between monomers
separated by a fixed contour distance.
Helices are favored only when the spatial geometry of the chain satisfies a
commensurability condition with this interaction spacing.
This route connects directly to classical helix--coil theories, such as those of
Gibbs--DiMarzio and Zimm--Bragg, but is reformulated here at the level of continuum
geometry and free energy.
A defining feature of Route~(B) is cooperativity: stabilization is not due to isolated
contacts but to runs of consecutively satisfied interactions, leading to sharp but
finite helix--coil transitions.
This mechanism is therefore particularly relevant to sequence-encoded polymers,
including DNA-like systems with strong hydrogen bonding.

An important outcome of the present theory is that both routes yield explicit and
quantitative stability criteria.
In Route~(A), helix formation is controlled primarily by the functional dependence on
the persistence length and the square of the thickness, together with the effective
cohesive strength.
In Route~(B), stability is governed by the persistence length, the sticker spacing,
the interaction strength, and the degree of cooperativity.
Using realistic parameter ranges, we have shown that Route~(A) can be marginal or
inaccessible in many generic bead--spring models—consistent with the observation of
globules, toroids, and rods but not helices—while Route~(B) naturally stabilizes
helices in biopolymers where strong periodic interactions are present.
In this sense, the theory provides not only positive predictions but also a diagnostic
interpretation of negative results.

More broadly, the present work clarifies that helices occupy a special position in the
polymer morphology landscape.
They are neither simple extensions of globules nor minor variants of toroids or rods.
Instead, they require additional geometric or energetic structure that is absent in
minimal collapse models.
By identifying these structures explicitly and placing them on a common free-energy
footing, we provide a systematic framework for understanding when helices should—and
should not—appear.

Finally, it is useful to place the present theoretical framework in the broader
context of continuum and field–theoretic descriptions that have proved
successful in soft condensed matter physics.  In particular, the analysis
developed here is closely related in spirit to the continuum approaches used in
the study of DNA condensation and electrostatic self–assembly, as discussed for
example in the perspective of Gelbart, Bruinsma, Pincus, and Parsegian
\cite{GelbartPT2000}.  In that work, the complex organization of DNA and other
polyelectrolytes was described using coarse–grained free–energy functionals in
which electrostatic interactions, elasticity, and confinement compete to select
ordered condensed structures such as toroids and cylinders.  The present theory
adopts a similar philosophy but focuses instead on the geometric and mechanical
ingredients controlling the morphology of a single semiflexible chain.  The
helical condensates identified here arise from the competition between bending
elasticity, cohesive interactions, steric constraints, and periodic contact
energetics, all expressed within a continuum free–energy framework.

At a deeper level, the theoretical structure also parallels the field–theoretic
approach developed in the theory of liquid crystals, where ordered structures
emerge from the minimization of coarse–grained free–energy functionals involving
orientational order parameters and elastic distortions
\cite{deGennesProst1993}.  In that language, helices represent states in which
curvature and torsion play roles analogous to elastic distortions of a director
field, while the spontaneous selection of left– or right–handed helicity
reflects a symmetry breaking similar to that encountered in chiral liquid–crystal
phases.  The present work therefore illustrates how ideas from polymer
statistical mechanics, electrostatic self–assembly, and liquid–crystal field
theory can be combined to construct a unified continuum description of
helical polymer condensates.

Several directions for future work are suggested by our results.
These include explicit simulation tests of the predicted helix windows in the relevant
parameter spaces, extensions to explicitly chiral interactions, and incorporation of
thermal fluctuations around the optimal helical geometry.
More generally, the framework developed here provides a bridge between classical
helix--coil theories and modern studies of polymer morphology, emphasizing that
helical order is a subtle and emergent phenomenon rather than a generic outcome of
polymer collapse.

%
\begin{center}
\textbf{Appendix A: Technical details of the packing/thickness mechanism (Route A)}
\end{center}

This appendix collects technical steps and asymptotic estimates underlying
Route~(A), which are not essential for the logical flow of the main text.
All defining equations, physical assumptions, and free-energy balances
associated with the packing/thickness mechanism have been presented explicitly
in Section~2.

\subsection*{A.1 Minimization of the nonlocal distance function}

The nonlocal steric constraint for a thick polymer requires that the distance
between non-neighboring contour points satisfy $d(\Delta s)\ge 2t$.
For a helical ansatz, the squared distance can be written in the dimensionless
form
\begin{equation}
\frac{d^2}{R^2}=g(x;u)=2(1-\cos x)+u^2 x^2,
\end{equation}
where $x=\Delta s/[R\sqrt{1+u^2}]$ and $u=P/(2\pi R)$.
The stationary points satisfy
\begin{equation}
g'(x;u)=2\sin x+2u^2 x=0.
\end{equation}
Besides the trivial minimum at $x=0$, a nontrivial minimum appears near
$x\simeq 2\pi$, corresponding to neighboring turns of the helix.

To obtain an analytic estimate, write $x=2\pi+\delta$ with $|\delta|\ll 1$.
Expanding $\sin(2\pi+\delta)\simeq \delta$ gives
\begin{equation}
\delta \simeq -\frac{2\pi u^2}{1+u^2}.
\end{equation}
Evaluating $g(x;u)$ at this stationary point yields
\begin{equation}
g_{\min}(u)\simeq 4\pi^2 u^2
\qquad (u \ \text{not too large}),
\end{equation}
so that the minimum distance scales as
\begin{equation}
d_{\min}\simeq 2\pi R u = P.
\end{equation}
This technical estimate justifies the statement made in the main text that
the closest approach of neighboring turns is controlled primarily by the pitch.

\subsection*{A.2 Remarks on thickness models}

The scaling results derived in Section~2 assume an effective tube thickness $t$.
The precise numerical prefactors depend on how thickness is implemented
(e.g.\ hard-core tube, effective excluded diameter, or three-body thickness
definitions used in ``thick polymer'' models).  These differences do not affect
the qualitative conclusion that the simultaneous enforcement of local curvature
and nonlocal self-avoidance selects finite values of $R$ and $P$ of order $t$.
%

\begin{center}
\textbf{Appendix B: Technical details of the periodic sticker mechanism (Route B)}
\end{center}

This appendix provides supporting derivations for Route~(B), which stabilizes
helices through periodic attractive interactions along the polymer backbone.
All defining equations, contact conditions, and free-energy expressions have
been given in Section~2.

\subsection*{B.1 Dimensionless form of the sticker contact condition}

For two monomers separated by $\Delta s=mb$, the squared distance along a helix
may be written in the dimensionless form
\begin{equation}
\frac{d^2(mb)}{(mb)^2}
=
\frac{2}{x^2(1+u^2)}\big(1-\cos x\big)
+
\frac{u^2}{1+u^2},
\end{equation}
where $x=mb/[R\sqrt{1+u^2}]$.
Setting $d(mb)=a_c$ yields the master constraint
\begin{equation}
\frac{a_c^2}{(mb)^2}
=
\frac{2}{x^2(1+u^2)}\big(1-\cos x\big)
+
\frac{u^2}{1+u^2},
\end{equation}
which implicitly determines the allowed region in $(R,u)$ space for which
periodic sticker contacts can be satisfied.

\subsection*{B.2 Commensurate limit and asymptotic simplifications}

In the commensurate regime $x\approx 2\pi n$ with integer $n$, the cosine term
vanishes and the contact condition simplifies considerably.  In this limit the
distance between sticker partners is dominated by axial separation, leading to
the inequality discussed in Section~2.
The resulting small-$u$ regime corresponds physically to helices with relatively
small pitch, which facilitate repeated contacts between distant backbone
segments.

\subsection*{B.3 Smooth-contact generalization}

In the main text we employed a sharp contact criterion for clarity.
More generally, one may replace the binary satisfaction factor
$\mathcal{P}(R,u)$ by a smooth function of $d(mb)$, representing a finite-range
attractive well.  This refinement removes discontinuities in the free energy
and allows standard minimization of the helix free-energy functional.
Such smoothing does not alter the scaling thresholds derived in Section~2.

%
%

%

\end{document}